\pdfoutput=1
\PassOptionsToPackage{pdfpagelabels=false}{hyperref}
\documentclass[fleqn,usenatbib,usedcolumn]{mnras}
\usepackage[british]{babel}             

\usepackage{newtxmath,newtxtext}

\usepackage[T1]{fontenc}

\usepackage{graphicx}    
\usepackage{amsmath}    
\usepackage{amssymb}    
\usepackage{etoolbox}
\makeatletter
\patchcmd\@combinedblfloats{\box\@outputbox}{\unvbox\@outputbox}{}{%
   \errmessage{\noexpand\@combinedblfloats could not be patched}%
}%
 \makeatother

\newcommand{\cfe}{\ensuremath{[\textrm{C}/\textrm{Fe}]}}
\newcommand{\nfe}{\ensuremath{[\textrm{N}/\textrm{Fe}]}}
\newcommand{\feh}{\ensuremath{[\textrm{Fe}/\textrm{H}]}}

\newcommand{\ebv}{\ensuremath{\textrm{E}(B-V)}}

\newcommand{\teff}{\ensuremath{\textrm{T}_\textrm{eff}}}
\newcommand{\logg}{\ensuremath{\log \textrm{g}}}
\newcommand{\eso}{ESO280-SC06}

\title[First observations of \eso]{The most metal-poor Galactic globular cluster: the first spectroscopic observations of \eso}

\author[J. D. Simpson]{Jeffrey D. Simpson$^{1}$\thanks{Email: \texttt{jeffrey.simpson@aao.gov.au}}\\
$^{1}$Australian Astronomical Observatory, 105 Delhi Rd, North Ryde, NSW 2113, Australia\\
}

\date{Accepted 2018 March 28. Received 2018 March 21; in original form 2018 January 8}

\pubyear{2016}

\begin{document}
\label{firstpage}
\pagerange{\pageref{firstpage}--\pageref{lastpage}}
\maketitle

\begin{abstract}
We present the first spectroscopic observations of the very metal-poor Milky Way globular cluster \eso. Using spectra acquired with the 2dF/AAOmega spectrograph on the Anglo-Australian Telescope, we have identified 13 members of the cluster, and estimate from their infrared calcium triplet lines that the cluster has a metallicity of $\feh=-2.48\substack{+0.06 \\ -0.11}$. This would make it the most metal-poor globular cluster known in the Milky Way. This result was verified with comparisons to three other metal-poor globular clusters that had been observed and analyzed in the same manner. We also present new photometry of the cluster from EFOSC2 and SkyMapper and confirm that the cluster is located $22.9\pm2.1$~kpc from the Sun and $15.2\pm2.1$~kpc from the Galactic centre, and has a radial velocity of $92.5\substack{+2.4 \\ -1.6}$~km\,s$^{-1}$. These new data finds the cluster to have a radius about half that previously estimated, and we find that the cluster has a dynamical mass of the cluster of $(12\pm2)\times10^3$~M$_\textrm{\sun}$. Unfortunately, we lack reliable proper motions to fully characterize its orbit about the Galaxy. Intriguingly, the photometry suggests that the cluster lacks a well-populated horizontal branch, something that has not been observed in a cluster so ancient or metal-poor.
\end{abstract}

\begin{keywords}
globular clusters: individual: \eso
\end{keywords}



\section{Introduction}\label{sec:intro}
The Milky Way Galaxy has 147 known globular clusters within its gravitational sphere of influence \citep[][2010 edition]{Harris1996}. {Almost\footnote{There are some ancient clusters that do not show evidence for multiple populations. e.g., Ruprecht 106 \citep{Villanova2013}.}} every globular cluster studied in detail with photometry and/or spectroscopy exhibits multiple populations, and although there are several proposed mechanisms for forming these populations during the infancy of clusters, none fully explain all the observed properties
 \cite[see the reviews of][and references therein]{Gratton2012b,Li2016b,Bastian2017}.
 
This family of stellar clusters continues to expand with deep imaging surveys \cite[e.g., discoveries by][]{Koposov2007,Kurtev2008,Balbinot2013,Koposov2015,Kim2016,Minniti2017a,Froebrich2017}, and already data from the space-based \textit{Gaia} astrometry mission has facilitated the detection of \textit{Gaia} 1 \& 2, two large, previously unknown clusters \citep{Koposov2017a,Simpson2017a,Mucciarelli2017a,Koch2017b}. \textit{Gaia} not only allows us to find over-densities on the two-dimensional plane of the sky, but also in the 5D spatial-dynamic space \citep{Andrews2017,Oh2016}.

While images of clusters and their derived photometry can tell us much about these clusters, it is through spectroscopy that we are able to measure radial velocities to confirm members, and infer stellar parameters and abundances of these stars. All these newly discovered clusters warrant spectroscopic investigation to understand their true nature --- whether they are truly co-eval, co-natal groups of stars, or perhaps simply statistical fluctuations in the star counts \citep[e.g., the false ``cluster'' Lod\'{e}n~1;][]{Han2016}.

However, these newly discovered clusters can be difficult to observe spectroscopically and have generally been poorly studied. This is because the stars of these clusters are usually faint, which can require large telescopes for high quality spectra; their faintness results in confusion with the field population along the line-of-sight; they have small apparent sizes, which makes it hard to multiplex the observations to efficiently use limited telescope time.

We have been undertaking an observing programme of unexplored clusters \citep{Simpson2017a, Simpson2016f}, and in this work, we continue this series by presenting the first spectroscopic results of the faint globular cluster \eso\ (18h09m06s, $-$46d25m24s). It has been the subject of only two papers that discussed the cluster in any detail \citep{Ortolani2000, Bonatto2008}. Although initially assumed to be an obscured open cluster when discovered in the ESO/Uppsala Southern Sky Survey \citep{Holmberg1977}, \citet{Ortolani2000} acquired images of the cluster and found its colour-magnitude diagram (CMD) morphology to be consistent with that of a metal-poor globular cluster, with only a small amount of reddening. Subsequently, \cite{Bonatto2008} used 2MASS photometry to estimate an overall absolute magnitude of $M_V\approx-4.9$, which would place it in the bottom 14\% of cluster luminosities.

The photometry of \citet{Ortolani2000} suggested that \eso\ had a sparse or perhaps non-existent horizontal branch (HB), and a poorly populated giant branch. Globular clusters are known to exhibit a wide range of morphologies of their HBs, and while age and metallicity clearly play an important role, there are one or more additional parameters involved; the so-called second parameter problem. This can be seen in "second-parameter" pairs of globular clusters \citep[e.g., NGC 288 \& NGC362;][]{Shetrone2000}, which have the same metallicity, but widely different HBs. Such morphological differences are especially marked at low metallicity. \citet{Milone2014} found that between clusters with metallicities $\feh\sim-2$ there was up to a 0.3 magnitude difference in the colour of their reddest HB stars. However, these results relies on a cluster having an observable HB and exclude clusters such as E3 \citep{Salinas2015}, AM-4 \citep{Carraro2009}, and Palomar~1 \citep{Sarajedini2007} which lack any obvious horizontal branch. {Where does \eso\ fit into this HB morphology distribution?}

This paper is structured as follows: in Section \ref{sec:reduction} we describe the observations and reduction of the spectra; Section \ref{sec:rv_members} explains the radial velocity measurement and member identification; Section \ref{sec:photometry} presents new CMDs of the cluster; in Section \ref{sec:metallicity} the metallicity of the cluster is estimated; Section \ref{sec:photom_feh} identifies possible unobserved members of the cluster using SkyMapper photometry; and Section \ref{sec:discussion} discusses these results, especially the very low metallicity and the lack of an obvious horizontal branch.

\section{Spectroscopic observations and data reduction}\label{sec:reduction}

\begin{figure}
    \includegraphics[width=\columnwidth]{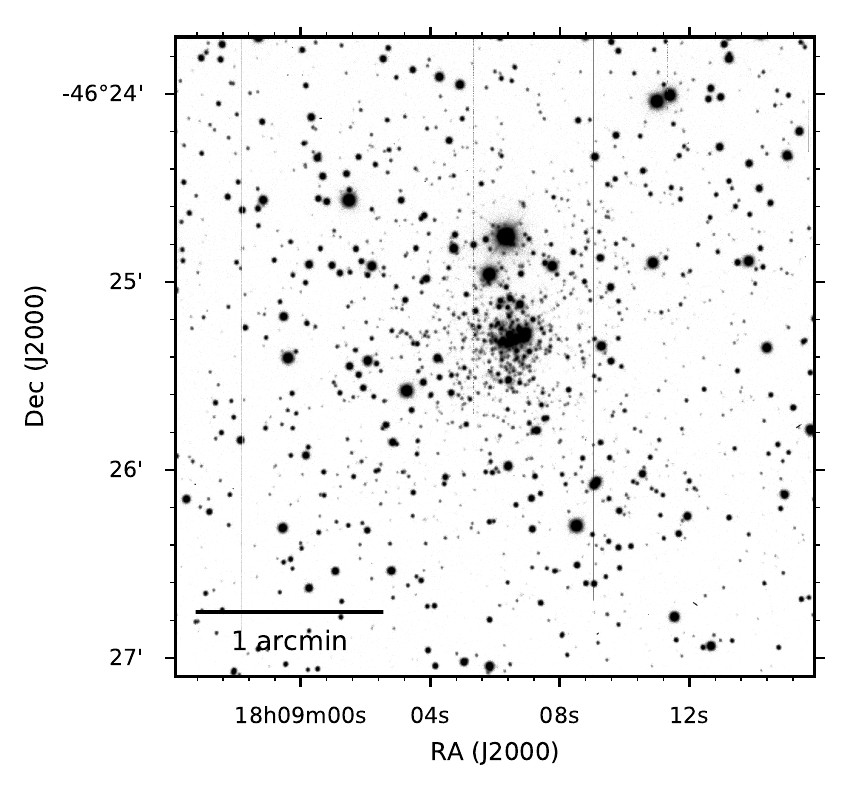}
    \caption{\eso\ as observed with a 100-second V filter exposure with the ESO Faint Object Spectrograph and Camera Version 2 on the ESO New Technology Telescope.}
    \label{fig:efosc}
\end{figure}

\begin{figure*}
    \includegraphics[width=\textwidth]{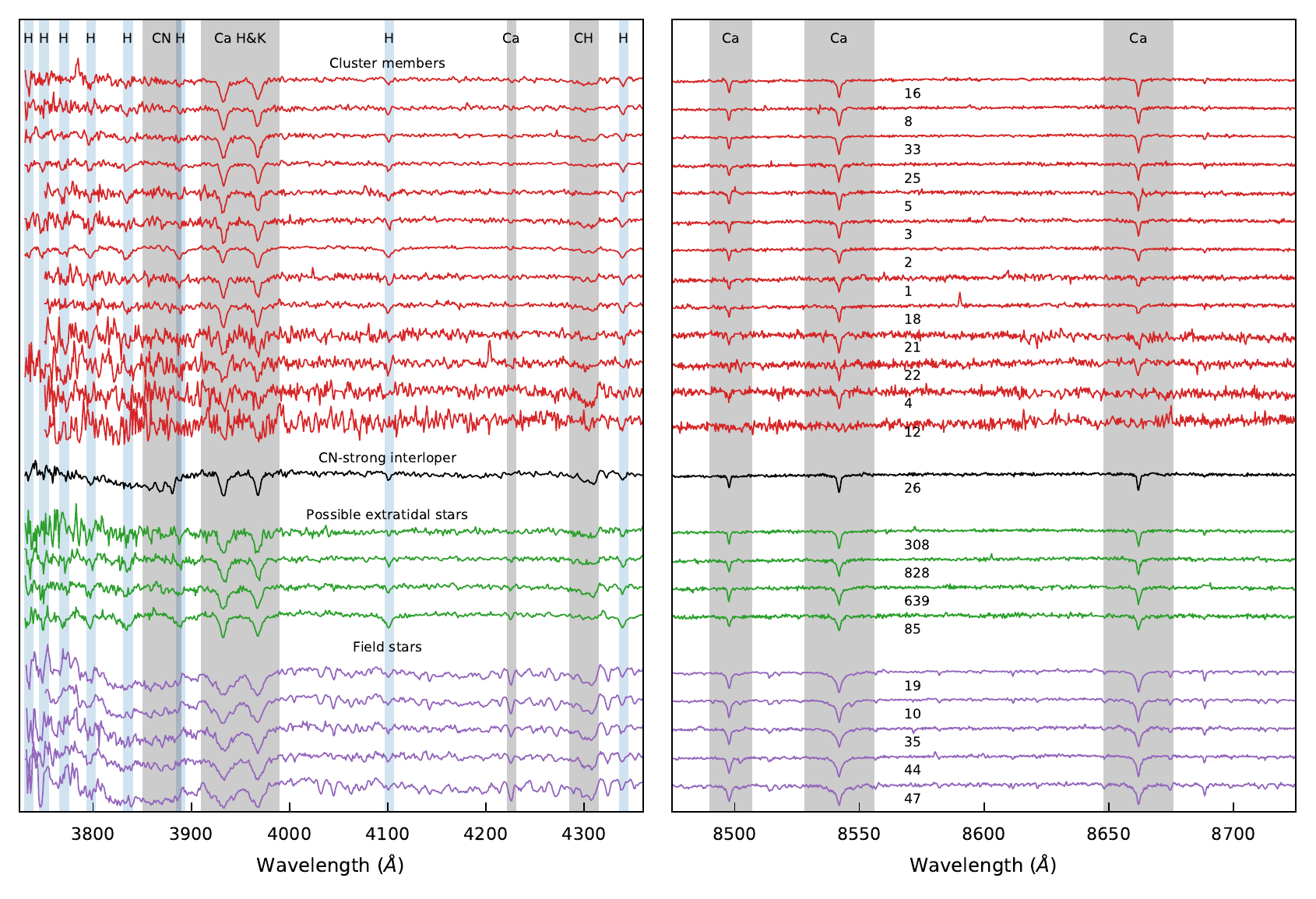}
    \caption{Portions of the blue camera (left) and red camera (right) spectra for four categories of stars observed: the 13 cluster member stars with `normal' spectra (red); one CN-strong interloper star that was at the cluster location and radial velocity but has a very strong CN band (black); the four possible extra-tidal stars (green); and five random giant field stars to contrast with the metal-poor cluster members (purple). In each category, the spectra are ordered by star brightness with the brightest star at the top (Table \ref{table:cluster_aaomega} presents the results in the same order). In the right panel, the number under each spectrum is the internally use star ID. The vertical shaded bands highlight spectral regions of interest. The member stars all have very weak-to-non-existent CN features, and only in some stars can the CH feature be seen. This is in contrast to the field stars which show strong CN and CH bands and also the strong Ca 4226~\AA\ line that is not present in the member stars due to their low metallicity.}
    \label{fig:spec_stacked}
\end{figure*}

\begin{table}
\caption{Details for the \eso\ observations with AAT/AAOmega. The signal-to-noise is per pixel for the red camera spectra.}
\label{table:aaomega_details}
\begin{tabular}{lrlllll}
\hline
Date & Total stars & Exp time (s) & G mag.\ range & SNR range \\
\hline
2016-07-09 & 355 & $4\times1200$ & 13.7--19.6 & 9--113 \\
2016-10-05 & 6 & $3\times1200$ & 15.2--17.9 & 16--64 \\
2016-10-05 & 6 & $3\times1200$ & 16.4--18.6 & 6--26 \\
2017-07-22 & 357 & $3\times1200$ & 15.4--18.0 & 3--58 \\
2017-07-22 & 351 & $3\times1200$ & 15.3--17.8 & 8--60 \\
\hline
\end{tabular}
\end{table}

\eso\ was observed over five nights in 2016 and 2017 (see Table \ref{table:aaomega_details} for observing details) with the 3.9-metre Anglo-Australian Telescope and its AAOmega spectrograph \citep{Sharp2006}, with the 392-fibre Two Degree Field (2dF) top-end \citep{Lewis2002}.

A total of 1075 stars were observed using five different 2dF plate configurations. Of these stars, 54 were within 3~arcmin of the cluster centre and were selected as they are near the cluster sequence on the colour-magnitude diagram (CMD). As shown in Figure \ref{fig:efosc}, the cluster has a small apparent size \citep[half light radius of 1~arcmin;][]{Bonatto2008} and the 2dF fibre buttons have a collision radius of 30--40~arcsec, hence the small number of possible cluster stars observed. The stars observed further than 3~arcmin from the cluster were selected with the aim of finding possible extra-tidal stars \citep[e.g.,][]{DaCosta2012,Navin2015,Navin2016,Simpson2017}. The plate configurations were created using the AAO's \textsc{configure} software \citep{Miszalski2006}, with 25 fibres per configuration assigned to sky positions.

AAOmega is a moderate resolution, dual-beam spectrograph. As in our previous cluster work \citep{Simpson2017,Simpson2016f}, the following gratings were used at their standard blaze angles: the blue 580V grating ($R\sim1200$; 3700--5800~\AA) and red 1700D grating ($R\sim10000$; 8340--8840~\AA). The 580V grating provides low-resolution coverage of the calcium H \& K lines and spectral regions dominated by CN and CH molecular features in cool giants. The 1700D grating was specifically designed to observe the near-infrared calcium triplet ($\sim8500$~\AA) at high resolution for precise radial velocity measurements. Also acquired were standard calibration exposures of a quartz lamp for defining the fibre traces on the raw images and He+CuAr+FeAr+CuNe arc lamps for wavelength calibration.

The raw images were reduced to 1D spectra using the AAO's \textsc{2dfdr} data reduction software \citep[][v6.46]{AAOSoftwareTeam2015} with the default \textsc{2dfdr} configuration appropriate for each grating. The standard spectral reduction steps were all performed automatically: bias subtraction using the overscan, determining the spectral traces on the raw images using the fibre flat, wavelength calibration using the arc exposure, extraction of the stellar spectra, sky subtraction using the fibres assigned to sky positions, and finally combining the individual exposures of each star. See Figure \ref{fig:spec_stacked} for examples of the reduced spectra.

\section{Membership identification}\label{sec:rv_members}

\begin{figure}
    \includegraphics[width=\columnwidth]{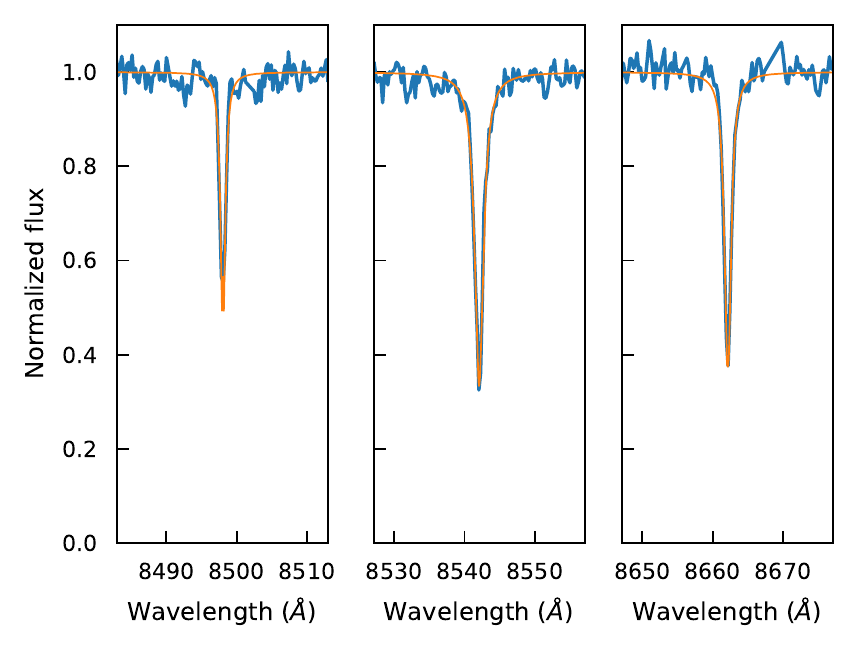}
    \caption{{Example of the Voigt functions (orange lines) fitted to the CaT lines of the normalized observed spectrum of Star 33 (blue lines).}}
    \label{fig:spec_fit}
\end{figure}

The radial velocities of the stars were measured from the reduced spectra using the near-infrared calcium triplet (CaT) lines at 8498.03, 8542.09 and 8662.14 \AA\ \citep{Edlen1956} using a method similar to that of \citet{Simpson2016f}. A number of realizations of each reduced spectrum were generated, with each realization having a random noise value added to each pixel point drawn from a Gaussian with a width equal to that pixel's noise value. Then for each realization, the following was performed.

The spectrum was shifted to zero barycentric velocity calculated by the \texttt{radial\_velocity\_correction} method of \textsc{astropy}. Any pixel whose noise value was larger than 1.5 times the median noise was masked out. Then with the strong CaT lines masked out as well, the spectrum was normalized by a four-degree Chebyshev polynomial with \textsc{scipy}'s \textsc{chebfit}. This normalization was refined with a straight line fit to the five continuum points defined by \citet{Carrera2013}. For each of the CaT line regions defined by \citet{Carrera2013}, a \texttt{Voigt1D} model \citep{McLean1994} from \textsc{astropy} was fitted with a Levenberg-Marquardt algorithm to the normalized spectrum {(Figure \ref{fig:spec_fit})}. From these fitted Voigts, the radial velocity and equivalent width of the three lines were calculated. The mean radial velocity of the three lines was found and the wavelength scale adjusted so that the spectrum was now at this rest wavelength. This whole procedure was then iterated for each spectrum realization until the velocity adjustment was $<1$~km\,s$^{-1}$. For a given realization, if any line had a radial velocity $>10$~km\,s$^{-1}$ from the median, it was ignored (and this particular realization would not be used to determine the equivalent width as such a radial velocity offset from the other lines would mean that line was poorly fitted, or missed entirely).

For each realization, the total equivalent width (EW) of the three lines was found, and the metallicity of the star calculated (see Section \ref{sec:metallicity}). The median, and half the range between the 16th and 84th percentiles were calculated for the RV, EW, and \feh\ for a given star from the N realizations (see Table \ref{table:eso280_members}).

\begin{table*}
\caption{Photometry and spectral results for the 13 probable members of \eso, the one CN-strong interloper, and the four possible extratidal stars. The ordering is the same as Figure \ref{fig:spec_stacked}. The full version of this table contains all $\sim1000$ stars observed. The photometry is sourced from: $G$ \textit{Gaia} DR1; $J,K_S$ 2MASS, and $v,g,z$ SkyMapper DR1.1.}
\label{table:eso280_members}
\begin{tabular}{rrrrrrrrrrrrr}
\hline
ID & RA & Dec & $G$ & $J$ & $K_S$ & $v$ & $g$ & $z$ & $v_r$ (km\,s$^{-1}$) & $\sum EW$ & \feh & r (arcmin) \\
\hline
\multicolumn{13}{c}{Cluster Members}\\
16 & 272.2590 & $-46.4149$ & $15.73$ & $14.17$ & $13.55$ & $17.57$ & $16.37$ & $15.33$ & $93.4\pm0.3$ & $3.1\pm0.1$ & $-2.45\pm0.03$ & 0.8 \\
8 & 272.2885 & $-46.4220$ & $15.85$ & $14.39$ & $13.82$ & $17.67$ & $16.52$ & $15.47$ & $91.1\pm0.4$ & $3.0\pm0.1$ & $-2.46\pm0.03$ & 0.6 \\
33 & 272.2742 & $-46.4504$ & $15.93$ & $14.34$ & $13.70$ & $17.91$ & $16.66$ & $15.47$ & $92.4\pm0.3$ & $3.2\pm0.1$ & $-2.39\pm0.03$ & 1.6 \\
25 & 272.2477 & $-46.4194$ & $16.27$ & $14.88$ & $14.20$ & $17.54$ & $16.84$ & $15.94$ & $92.3\pm0.5$ & $2.7\pm0.1$ & $-2.55\pm0.04$ & 1.2 \\
5 & 272.2780 & $-46.4183$ & $16.37$ & $14.95$ & $14.05$ & $$ & $$ & $$ & $93.9\pm0.6$ & $2.9\pm0.1$ & $-2.49\pm0.05$ & 0.3 \\
3 & 272.2674 & $-46.4231$ & $16.45$ & $14.92$ & $14.29$ & $$ & $17.10$ & $16.09$ & $94.7\pm0.8$ & $2.6\pm0.1$ & $-2.59\pm0.05$ & 0.3 \\
2 & 272.2767 & $-46.4210$ & $16.56$ &  &  & $$ & $$ & $$ & $96.2\pm0.6$ & $2.6\pm0.1$ &  & 0.2 \\
1 & 272.2765 & $-46.4250$ & $17.09$ & $15.49$ & $15.00$ & $$ & $$ & $$ & $96.1\pm1.6$ & $2.3\pm0.3$ & $-2.58\pm0.16$ & 0.1 \\
18 & 272.2797 & $-46.4382$ & $17.23$ & $15.82$ & $14.99$ & $$ & $17.99$ & $16.96$ & $92.5\pm0.9$ & $2.6\pm0.1$ & $-2.47\pm0.07$ & 0.9 \\
21 & 272.2627 & $-46.4091$ & $17.59$ & $16.38$ & $15.33$ & $$ & $18.10$ & $17.38$ & $92.1\pm2.5$ & $2.9\pm1.6$ & $-2.21\pm0.18$ & 1.0 \\
22 & 272.2908 & $-46.4366$ & $17.67$ & $16.35$ & $15.43$ & $$ & $18.42$ & $17.33$ & $88.2\pm2.1$ & $1.8\pm0.5$ & $-2.78\pm0.27$ & 1.0 \\
4 & 272.2677 & $-46.4248$ & $18.33$ &  &  & $$ & $$ & $$ & $94.8\pm3.4$ & $2.6\pm0.9$ &  & 0.3 \\
12 & 272.2884 & $-46.4306$ & $18.49$ &  &  & $$ & $$ & $$ & $86.8\pm19.8$ & $2.3\pm0.9$ &  & 0.7 \\
\multicolumn{13}{c}{CN-strong interloper}\\
26 & 272.2615 & $-46.4419$ & $16.38$ & $14.81$ & $14.26$ & $$ & $17.06$ & $15.97$ & $91.5\pm0.5$ & $3.0\pm0.1$ & $-2.39\pm0.03$ & 1.2 \\
\multicolumn{13}{c}{Possible extratidal stars}\\
308 & 272.8380 & $-46.6799$ & $16.13$ & $14.64$ & $13.90$ & $17.96$ & $16.81$ & $15.76$ & $84.6\pm0.6$ & $3.2\pm0.1$ & $-2.36\pm0.04$ & 27.9 \\
828 & 271.0598 & $-46.5165$ & $16.47$ & $15.07$ & $14.33$ & $18.19$ & $17.11$ & $16.14$ & $85.6\pm0.5$ & $3.4\pm0.1$ & $-2.19\pm0.06$ & 50.5 \\
639 & 271.4377 & $-45.9871$ & $16.94$ & $15.44$ & $14.76$ & $$ & $17.49$ & $16.55$ & $86.5\pm1.2$ & $3.4\pm0.1$ & $-2.11\pm0.06$ & 43.5 \\
85 & 272.2522 & $-46.5361$ & $16.98$ & $15.91$ & $14.70$ & $$ & $$ & $$ & $96.7\pm1.2$ & $3.7\pm0.2$ & $-1.99\pm0.10$ & 6.8 \\
\hline
\end{tabular}
\end{table*}

\begin{figure}
    \includegraphics[width=\columnwidth]{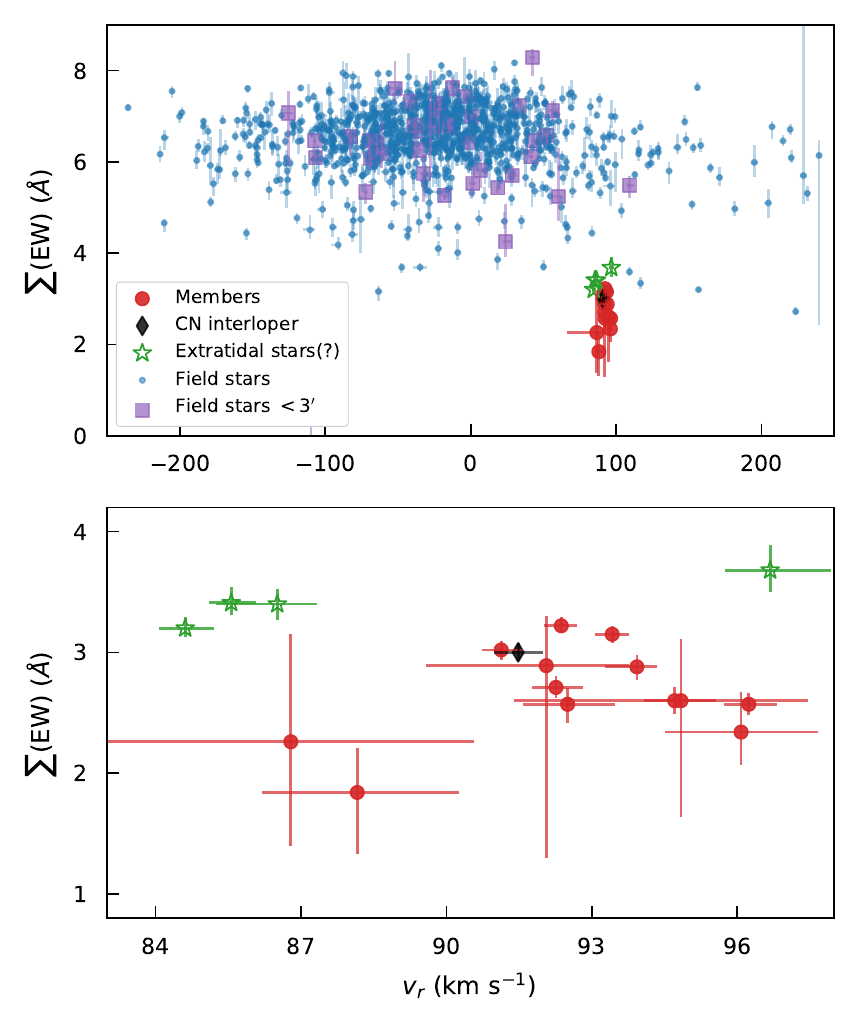}
    \caption{The members of the cluster stand out with their common radial velocities and weak calcium triplet line strengths. The top panel shows all the stars observed and the bottom panel shows only those stars with $\sum\mathrm{EW}<4$~\AA\ and $84<v_r<98$~km\,s$^{-1}$. Fourteen stars were initially identified as possible cluster members. But the inspection of the spectra (Figure \ref{fig:spec_stacked}) one had a much stronger CN band than the other stars and was likely not a cluster star. There are four potential extra-tidal stars with similar radial velocities as the cluster stars.}
    \label{fig:ew_rv}
\end{figure}

Of the 54 stars observed within 3~arcmin of the cluster, 14 stars had radial velocities in the range $85<v_r<98$~km\,s$^{-1}$ and with $\sum\mathrm{EW}_\mathrm{CaT}<4$~\AA\ (Figure \ref{fig:ew_rv}). These stars are preliminarily classified as possible cluster members. Star 12 has the lowest radial velocity of these 14 stars, but it had very low signal spectra ($\mathrm{SNR}=6$ per pixel in the red camera spectrum), so its radial velocity is much more uncertain the rest of the possible members. The bulk of the stars observed (outside the 3~arcmin radius) had total equivalent widths $>\sim4$~\AA\ and radial velocities within a range centred on $-20$~km\,s$^{-1}$ with a standard deviation of $\pm76$~km\,s$^{-1}$. But there are four stars at similar velocities ($v_r=84,86,87,97$~km\,s$^{-1}$) to the potential cluster members: these are classified as possible extra-tidal stars.

In Figure \ref{fig:spec_stacked}, a portion of the blue and red spectra of the 14 possible member stars are plotted, as well as the four extra-tidal stars, and for comparison, five random field giant stars. The blue camera regions plotted contain the calcium H \& K lines, the prominent CN \& CH molecular features and most of the hydrogen Balmer lines; while the red camera spectra portion plotted contains the CaT lines. The 14 possible members are divided into two categories based upon visual inspection of their spectrum: 13 stars with ``normal'' spectra; and one star with a strong CN band. Three of the four extratidal stars show strong Balmer lines. In Sections \ref{sec:excluded} and \ref{sec:extratidal} the morphology of the spectra is discussed and we exclude the CN-strong star, leaving the 13 likely members.

\subsection{Excluded ``members''}\label{sec:excluded}

Of the 14 possible cluster members, one stand out with different spectra from the rest, with a strong CN band (Figure \ref{fig:spec_stacked}). The CN-strong star is peculiar as none of the other cluster stars exhibits such a strong CN bandhead.  The spectrum lacks the strong C$_2$ features that would be found in a very carbon-enhanced star \citep[e.g.,][]{Sharina2012}. Instead, it looks like a typical CN-strong star observed in other clusters \cite[e.g.,][]{Norris1981, Simpson2017}. It is not unusual for clusters to show a range of nitrogen abundances, but at the low metallicity of \eso\ ($\feh\approx-2.5$; see Section \ref{sec:metallicity}), normal levels of carbon and nitrogen variations would not make an observable difference in the CN band strengths \citep{Shetrone2010}.

\begin{figure}
    \includegraphics[width=\columnwidth]{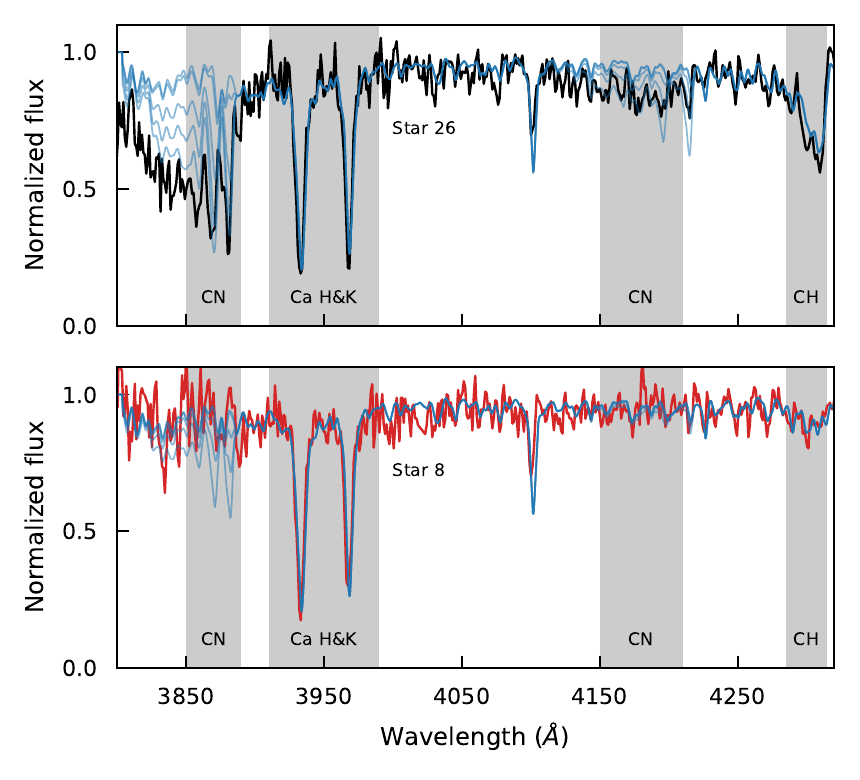}
    \caption{Observed and synthetic spectra for the CN-strong star (Star 26; top panel) and a CN-normal star of the same $J-K_S$ colour (Star 8; bottom). For both panels the synthetic spectra using a model atmosphere of $\teff=5000$~K, $\logg=1.5$, $\feh=-2.5$ and $\nfe=0.0, 1.0 ,2.0 ,2.5, 3.0$. For the top panel $\cfe=+0.5$ (to fit the strength of the CH band) and the bottom panel $\cfe=-0.5$. In order to match the CN band strengths of the CN-strong star, it would be necessary to have $\nfe\sim3$.}
    \label{fig:spec_syn}
\end{figure}

Spectrum synthesis was used to estimate the nitrogen abundance of this CN-strong star and determine if it was within the reasonable range for a globular cluster. Spectra were synthesized with \textsc{moog} \citep[][2009 Version]{Sneden1973} as implemented in \textsc{iSpec} \citep{Blanco-Cuaresma2014} using the line list extracted from the VALD database \citep{Kupka2011} provided by \textsc{iSpec}. MARCS stellar atmosphere models \citep{Gustafsson2008} for  one solar-mass giants with spherical geometry, solar-scaled composition, and microturbulent velocity of $v_t=2.0$~km\,s$^{-1}$ were used. The 2MASS photometry of the star (see Section \ref{sec:photometry} and Table \ref{table:eso280_members}) provides an estimated effective temperature of $\teff\approx5000$~K \citep{GonzalezHernandez2009} and for this temperature the isochrones in Section \ref{sec:photometry} give a surface gravity of $\logg\approx1.5$. It should be noted that these syntheses have not been done for precise abundance determination so these parameters are estimates.

In Figure \ref{fig:spec_syn} is the comparison of the spectra of two stars with the same $J-K_S$ colour (and therefore temperature): the CN-strong star (Star 26) and a CN-normal star (Star 8). Not only does Star 26 have strong CN bands, it has a strong CH band. For a metallicity of $\feh=-2.5$ it is necessary to have $\cfe\approx+0.5$ and $\nfe\sim+3$ to match the strength of the CH and CN bands of this star. Whereas, for the comparison CN-normal star, it is estimated that $\cfe=-0.5$ and it is only at extreme nitrogen enhancement that we see any change in the synthesized CN band strength.

Super-solar carbon abundances are not typical in metal-poor globular clusters: M92 \citep{Bellman2001,Roederer2011}, NGC5024 and NGC5466 \citep{Shetrone2010,Lamb2015} have all been found to have $\cfe<0$, and their nitrogen abundances are much lower than our estimate for star Star 26, with $\nfe<2$. The stars of the metal-poor cluster M5 have mostly $\cfe<0$, though with a handful of stars up to $\cfe\sim+0.5$, but all have $\nfe<2$ \citep{Cohen2002}. Of course, the carbon-nitrogen anticorrelation of globular clusters means that it is the carbon-poor stars that are nitrogen-rich. For example, in M5, the most nitrogen-rich stars have $\cfe<-0.5$.

The extreme nitrogen abundance combined with the high carbon abundance leads to the conclusion that this CN-strong star is simply a serendipitous interloper along the line-of-sight instead of being a real member of the cluster. If it is closer to the Sun than \eso\ it will be less luminous than assumed here, and therefore will have a higher metallicity as determined from the equivalent widths of the calcium triplet lines (the effect of distance is also discussed in Section \ref{sec:metallicity} in the context of errors in the distance modulus). A more metal rich star would not need to be as enhanced in nitrogen as we estimate above.


\subsection{Extra-tidal stars?}\label{sec:extratidal}

\begin{figure}
    \includegraphics[width=\columnwidth]{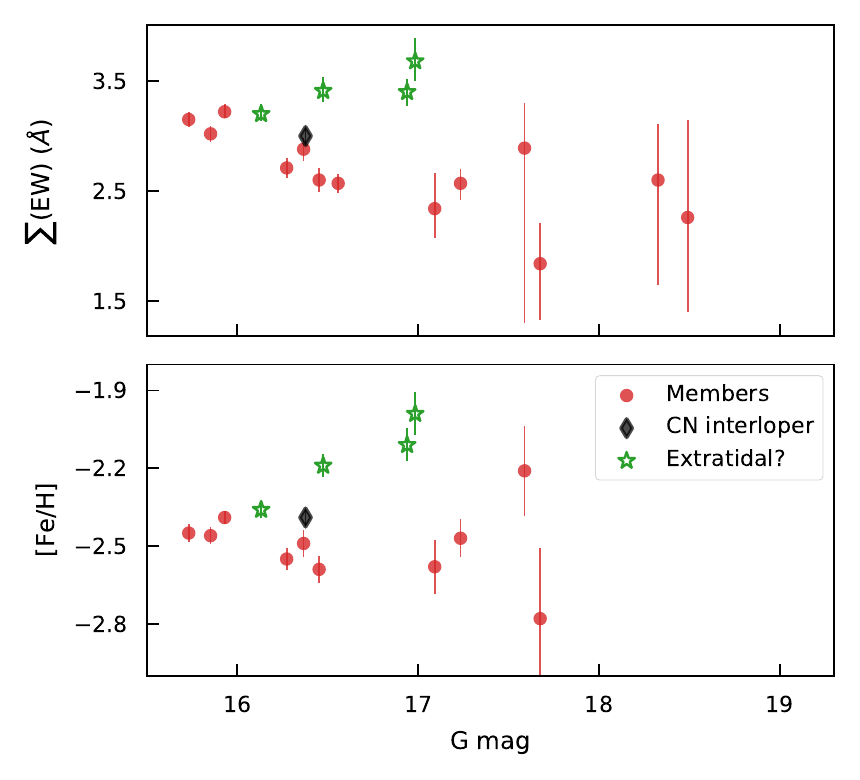}
    \caption{For stars in a globular cluster on the RGB  the CaT EW should be correlated with the luminosity of the star. In the top panel the magnitude of the stars are plotted against its CaT EW sum, and in the bottom panel it is against the metallicity calculated in Section \ref{sec:metallicity}. We find that the three of the four possible extra-tidal stars do not follow the expected trend, and conclude that they are low metallicity F-type field stars. Some of the stars are missing from the bottom panel because they lacked 2MASS $K_S$ magnitudes.}
    \label{fig:phot_ew_feh}
\end{figure}

The spectra (Figure \ref{fig:spec_stacked}) of three of the four extratidal stars show strong Balmer features {as would be expected for a foreground turn-off star.} The photometry of these possible extratidal stars can be used to exclude these three of the stars from being related to the cluster. For stars which at the same distance and with the same metallicity --- i.e., those in a globular cluster --- the equivalent widths of the CaT lines should be correlated with their magnitude, with the coolest (and therefore brightest) stars having the strongest lines. All four extra-tidal stars have large CaT EWs (Figure \ref{fig:ew_rv}), but they are not the brightest stars observed, and for three of the four, their EWs do not follow the expected trend with luminosity (Figure \ref{fig:phot_ew_feh}). The brightest possible extra-tidal star (Star 308) does have EW and photometry consistent with it being a lost cluster member. On the spectra plot (Figure \ref{fig:spec_stacked}) it is the top extra-tidal star plotted, and its (noisy) spectrum is much less dominated by the Balmer features. High-resolution follow-up is required to chemically tag the star to the cluster members.

\section{A new colour-magnitude diagram for \eso}\label{sec:photometry}

Three large photometric surveys include \eso\ in their footprint, and have catalogued most of the potential members: the SkyMapper Shallow Survey DR1.1 \citep[][\textit{uvgriz}; and see also Section \ref{sec:photom_feh}]{Wolf2017}, the VISTA Hemisphere Survey DR4.1 \citep[VHS;][$J$, $K_S$]{McMahon2013}, and the Two Micron All-Sky Survey \citep[2MASS;][$J$, $H$, $K_S$]{Skrutskie2006}. Unfortunately, in all cases, their faint limits were too shallow to reach the turn-off magnitude as found by \citet{Ortolani2000}. There is also confusion between the field population and the cluster so that the cluster sequence is not obvious on CMDs produced from their photometry.

Fortunately, new images of \eso\ were acquired on the night of 2015 May 17 with the ESO Faint Object Spectrograph and Camera Version 2 \citep[EFOSC2;][]{Buzzoni1984, Snodgrass2008} on the 3.58-metre New Technology Telescope (NTT) using the V\#641 (Bessell V)  and I\#705 (Gunn i) filters (V image shown in Figure \ref{fig:efosc}). These observations along with their calibration frames were retrieved from the ESO archive and were reduced using \textsc{esorex} with the EFOSC2 pipeline recipes (v2.2.5). Crowded field, point-spread function photometry was performed on each image using Iraf's \textsc{daophot}/\textsc{allstar} \citep{Stetson1987} to measure the instrumental magnitudes of the stars. For calibrations we used stars from \citet{Landolt1983,Landolt1992}. There are no very metal-poor isochrones available in Bessell $V$ and Gunn $i$ magnitudes, so it was necessary to linearly transform to other systems, in this case, UCAC $V$ and SkyMapper $i$. The SVO Filter Profile Service \citep{Rodrigo2017, Rodrigo2017a} was used to identify these as the most similar filters with available isochrones. There were only five stars in common between the EFOSC2 observations and UCAC4 with V magnitudes, so an additive offset of 0.5 magnitudes was applied. There were 63 stars in common with SkyMapper, with the transformation being $i_\textrm{SM}=1.0031304\times i_\textrm{Gunn}+0.36811$.


\begin{figure}
    \includegraphics[width=\columnwidth]{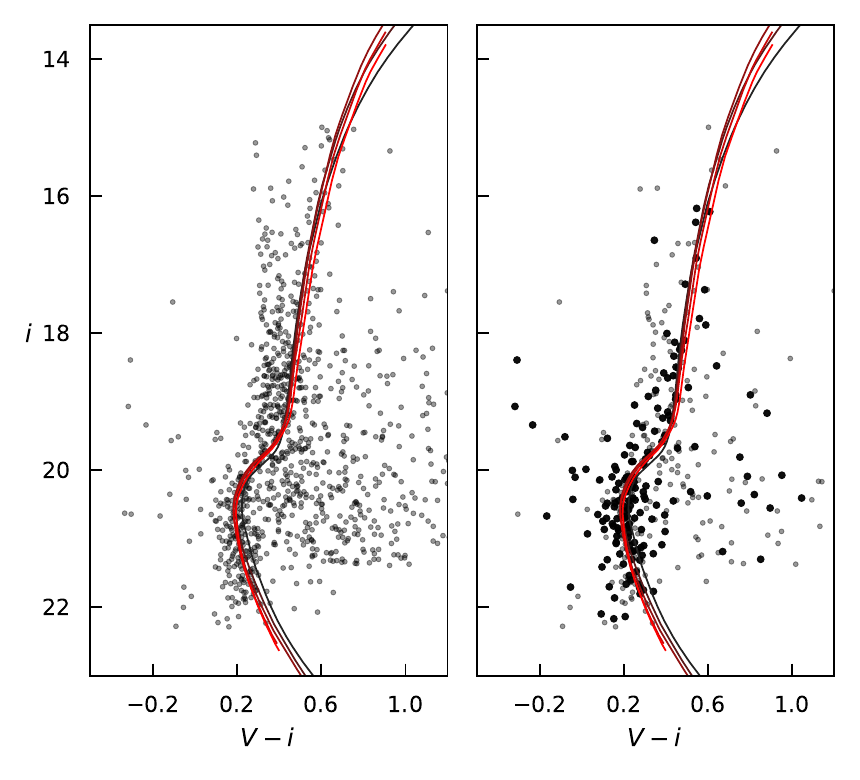}
    \caption{Colour-magnitude diagram of \eso\ produced from the EFOSC2 images of the cluster. The left panel shows all the stars in the full frame of Figure \ref{fig:efosc}, and the right panel shows only those stars within 60~arcsec (small dots) and 30~arcsec of the cluster centre (larger dots). The cluster sequence has been fitted with 13-Gyr Dartmouth isochrones for metallicities of $\feh=(-1.5,-2.0,-2.5,-3.0,-3.5)$.}
    \label{fig:initial_cmd}
\end{figure}

The left panel of Figure \ref{fig:initial_cmd} shows the CMD created from the EFOSC2 photometry. This photometry extends to fainter magnitudes than that of \citet{Ortolani2000}, and although many of the stars across the full field are Galactic disk dwarfs along the line-of-sight, there is a clear cluster sequence. Focusing on only those stars close to the cluster (right panel of Figure \ref{fig:initial_cmd}) shows that there are an obvious turn-off and giant branch, and the turn-off morphology is consistent with an old metal-poor cluster.

We choose 13-Gyr Dartmouth isochrones with $\feh=(-1.5,-2.0,-2.5,-3.0,-3.5)$ and $[\alpha/\mathrm{Fe}]=+0.4$, which were fitted by eye to to the cluster sequence visible on the EFOSC2 CMD. Unfortunately, there are no isochrones transformed to SkyMapper photometry that include the HB and AGB phases, but there is no obvious HB, so this does not affect our results. All else remaining the same, the turn-off gets fainter with decreasing metallicity, such that there is a small range of possible distance moduli and reddenings for the cluster. For the metallicity range $-1.5>\feh>-3.5$ we find $16.6<(m-M)_0<17.0$ and $0.02<\ebv<0.08$. The only literature estimate for the cluster are found in \citet{Ortolani2000} --- $(m-M)_0=16.7;\ebv=0.07$ --- so our new values do not represent a departure. The reddening is smaller than that estimated from the all-sky reddening maps determined by \citet{Schlegel1998} and \citet{Schlafly2011}, who reported $\ebv=0.16$ and $0.13$ respectively for the location of \eso.

With a metallicity of $\feh=-2.5$ (Section \ref{sec:metallicity}), we adopt a distance modulus of $16.8\pm0.2$, which places the cluster $22.9\pm2.1$~kpc from the Sun and $15.3\pm2.1$~kpc from the Galactic centre. This is equivalent to Cartesian coordinates centred on the Galactic centre of $[X,Y,Z]=[13.5\pm2.0,-5.1\pm0.5,-5.0\pm0.5]$~kpc (using the default solar position adopted by \textsc{astropy} v3.0\footnote{Specifically, the Galactic Centre position of $(\mathrm{RA},\mathrm{Dec})=(266.4051,-28.936175)$ \citep{Reid2004}, the Galactocentric distance of the Sun of 8.3~kpc \citep{Gillessen2009}, and height of the Sun above the Galactic midplane is taken to be 27 pc \citep{Chen2001a}}). The cluster is located on the edge of the Galactic disk and potentially in the halo \citep{Bland-Hawthorn2016}. None of the cluster members are bright enough for the Tycho-Gaia astrometric solution \citep{Michalik2014}, nor have reliable proper motions from UCAC5 \citep{Zacharias2017} or HSOY \citep{Altmann2017} so it is not possible to compute an orbit for \eso\ and comment further on its motion about the Galaxy.

\begin{figure}
    \includegraphics[width=\columnwidth]{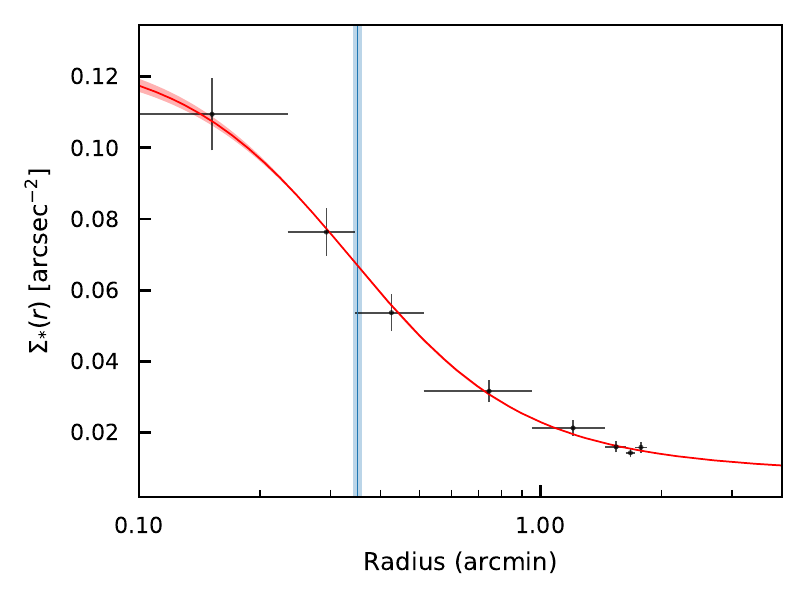}
    \caption{{Stellar density profile (SDP) of \eso\ derived from EFOSC2 images. The black dots are the star counts in each annulus, with the horizontal error bars defining the inner and outer radii of the annulus. The vertical bars are the standard deviation of the star counts for the four sectors in each annulus. The solid red line is the best fitting King profile and the red shaded region shows the $1\sigma$ confidence intervals. The blue lines and blue shaded regions indicate the locations of the $r_0$.}}
    \label{fig:sdp}
\end{figure}

{In \cite{Bonatto2008} structural parameters for \eso\ were found from King-like profile fitting to 2MASS-derived stellar density profiles. With the new deeper imagery from EFOSC2 we can determine updated structural parameters using the modified method of \citet{Miocchi2013} presented in \citet{Simpson2016f}. Briefly, the stellar density is found in annuli of varying radii around the centre of the cluster and these density values are fitted by a King profile pre-computed by \citet{Miocchi2013}. These profiles have a characteristic scale length $r_0$, which is similar in value, but different to, the core radius $r_c$. \citet{Miocchi2013} also determined the limiting radius $r_l$ for their profiles, which is the radius at which the projected density goes to zero. They define the concentration parameter $c\equiv\log(r_l/r_0)$. The stellar density profile is shown in Figure \ref{fig:sdp} and for \eso\ we have determined that $r_0 = 21.0\pm0.6$~arcsec, and that the cluster has a limiting radius $r_l= 170^{+30}_{-110}$~arcmin and a concentration $c = 2.7^{+0.1}_{-0.5}$. The limiting radius is very uncertain because of the small field-of-view of the EFOSC2 images, which likely does not extend to the tidal radius of the cluster. The $r_0$ is about half of that found by \cite{Bonatto2008}. With a heliocentric distance of $22.9\pm2.1$~kpc, \eso\ has a radius of $r_0=2.3\pm0.2$~pc, which is typical of Milky Way or M31 globular clusters \citep[e.g.,][]{VanDenBergh2010}.}

\begin{figure*}
    \includegraphics[width=\textwidth]{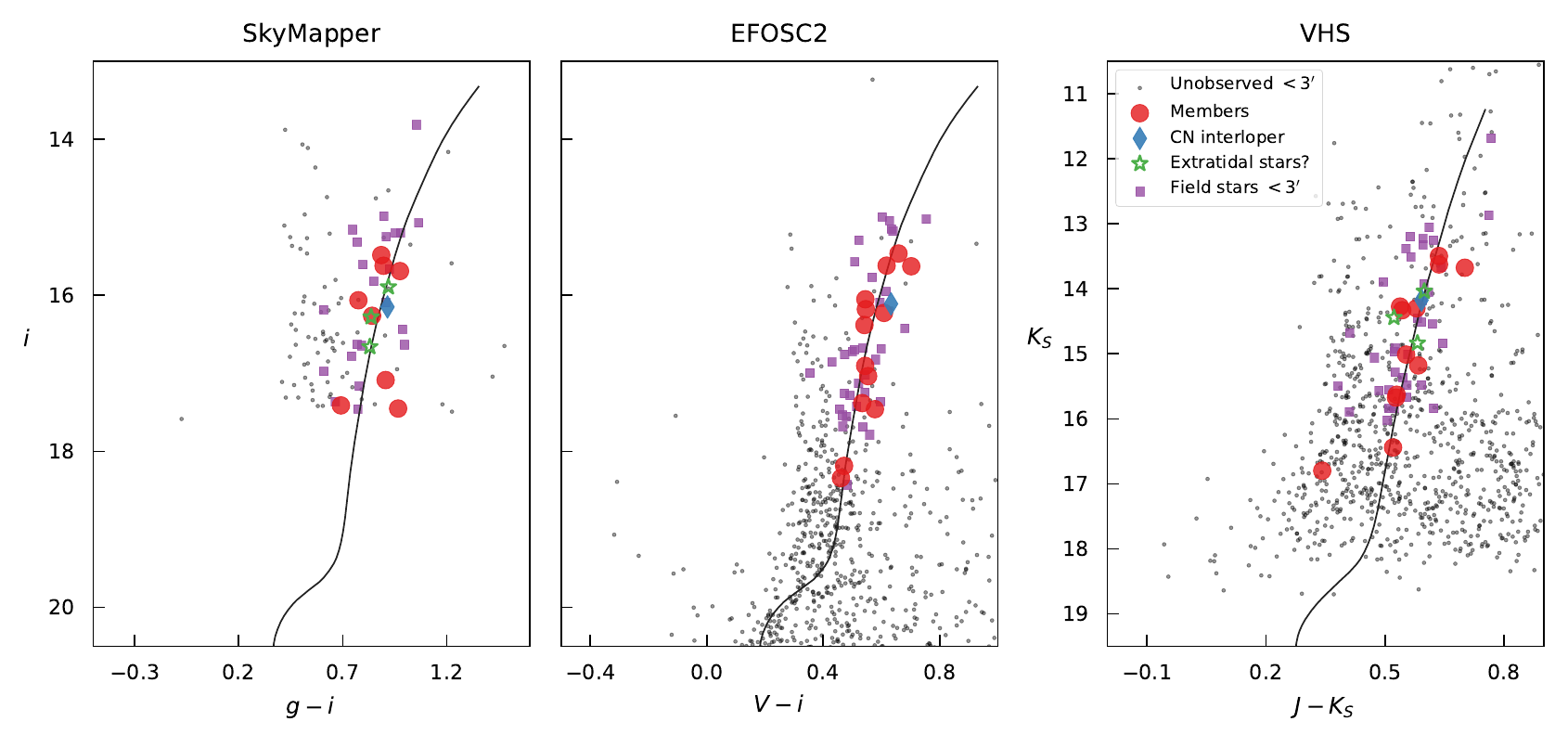}
    \caption{Three views of the CMD of \eso. From left-to-right is photometry from SkyMapper, EFOSC2, {and VHS}. The grey dots are all stars within 3~arcmin that were unobserved. The 13-Gyr isochrone of $\feh=-2.5$ has been fitted by eye to the EFOSC2 photometry and then assuming appropriate photometric transformations, placed on the other photometry. Two things are evident: 1) there are few unobserved bright giants; 2) the photometric catalogues suggest there is not a well-populated horizontal branch.}
    \label{fig:cmds}
\end{figure*}

The identified cluster members, possible extra-tidal stars, and known field stars are placed in context of the colour-magnitude diagrams created from the photometry of SkyMapper, {VHS,} and EFOSC2 in Figure \ref{fig:cmds}. For the SkyMapper photometry, in order acquire a clean dataset of reliable sources, it was required that $\mathrm{\texttt{class\_star}}>0.9$, $\mathrm{\texttt{nch\_max}} = 1$, $\mathrm{\texttt{flags}}=0$, and $\mathrm{\texttt{nimaflags}}=0$. {For the VHS photometry, it was required that $\mathrm{\texttt{PSTAR}}>0.99$ and $\mathrm{\texttt{PNOISE}}<0.01$.}

With no obvious horizontal branch on the CMDs, it was necessary to use the turn-off region visible in the EFOSC2 photometry as the anchor for the distance modulus and reddening of the cluster. To place the isochrones on the other CMDs, the distance modulus and reddening were transformed to the other bands using the following bandpass absorption coefficients: $R_g=2.986,R_i=1.588,R_z=1.206,R_J=0.723,R_K=0.310$ \citep{Schlegel1998,Wolf2017}. The isochrone fits show that \eso\ is metal-poor, with $\feh<-1.5$, but the available colours are not particularly metallicity-sensitive for this age and metallicity regime. In Section \ref{sec:metallicity} the metallicity of the cluster is estimated from the near infrared calcium triplet lines.

The most striking feature of the CMDs is the lack of any obvious horizontal branch. \citet{Ortolani2000} state there are four horizontal branch stars in their sample but we are uncertain if there is any HB visible on these CMDs. The region of the CMD {where we might find a red HB} coincides with the field dwarf sequence. We did spectroscopically observe several stars in this region of the CMD and found them to be field stars, but there are about ten unobserved possible red HB stars and they should be a high priority for future observations of the cluster.

%

A gravity sensitivity colour index could be created from the SkyMapper photometry, such as $u-v-0.2(g-i)$ \citep{Keller2007, Akhter2013, Wolf2017}. This can distinguish main sequence stars from blue horizontal branch stars of the same colour. Unfortunately, none of the aforementioned possible HB stars have $u$ and $v$ photometry, so we are unable to disentangle stars in this way.

\section{Cluster metallicity}\label{sec:metallicity}

\begin{figure}
    \includegraphics[width=\columnwidth]{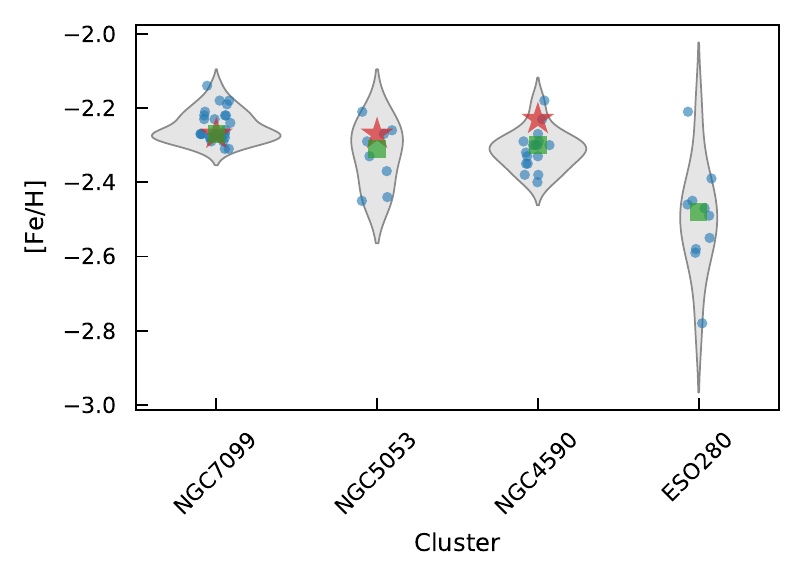}
    \caption{The metallicities determined from the CaT lines of stars for \eso\ and three comparison metal-poor clusters. The small blue dots are the individual cluster member metallicities, with their univariate kernel density estimate in grey. The large red star is the literature metallicity of the cluster \citep{Harris1996} and the large green square is the median metallicity found in this work. The difference between this work's metallicity estimates and the literature value is $<0.07$~dex. The two most outliers for \eso\ are the faintest stars which had the most uncertain metallicities.}
    \label{fig:violin}
\end{figure}

\begin{table*}
\caption{Radial velocities and metallicities determined by this work for \eso\ and three comparison clusters. For a given cluster, the radial velocity and metallicity values are the median of its member stars with the 16th and 84th percentile ranges. Literature values are taken from \citet[][H96]{Harris1996}. For \eso\ only 10 of the 13 members identified had \feh\ estimates because two of the stars lacked the required $K_S$ photometry.}
\label{table:cluster_aaomega}
\begin{tabular}{llrrrrrr}
\hline
Cluster & Stars & $v_r$ (km\,s$^{-1}$) & H96 $v_r$ (km\,s$^{-1}$) & \feh & H96 \feh & $\Delta\feh$ \\
\hline
ESO280 & 10& $92.5\substack{+2.4 \\ -1.6}$ &  & $-2.48\substack{+0.06 \\ -0.11}$ &  &  \\
NGC4590 & 17& $-91.9\substack{+1.4 \\ -2.1}$ & $-94.7\pm2.5$ & $-2.30\substack{+0.02 \\ -0.06}$ & $-2.23$ & $-0.07$ \\
NGC5053 & 8& $44.2\substack{+1.8 \\ -1.9}$ & $44.0\pm1.4$ & $-2.31\substack{+0.05 \\ -0.12}$ & $-2.27$ & $-0.04$ \\
NGC7099 & 27& $-184.1\substack{+1.6 \\ -2.3}$ & $-184.2\pm5.5$ & $-2.27\substack{+0.06 \\ -0.01}$ & $-2.27$ & $0.00$ \\
\hline
\end{tabular}
\end{table*}

One of the fundamental parameters of a globular cluster is the metallicity of its constituent stars. The acquired AAOmega spectra cannot be used to determine the metallicity of the stars using classical line-by-line equivalent width measurements of neutral and ionized iron lines. But the red camera spectra contains the near infrared calcium triplet (CaT) lines, which have been used extensively for estimating the metallicity of stars \cite[e.g.,][]{Simpson2016f, Simpson2017a}. As in our previous works, the relationship from \citet{Carrera2013} was used to relate the luminosity of a star and the equivalent widths of its CaT lines to the metallicity of the star. An important feature of the \citet{Carrera2013} method is that it is calibrated with very metal-poor stars, down to metallicities of $\feh=-4$.

The overlap of the available photometry of \eso\ members and the photometric systems used by \citet{Carrera2013} limits us to using the absolute 2MASS\footnote{{As shown in Section \ref{sec:photometry} there is VHS $K_S$ photometry of the cluster, but this is not available for all of the comparison clusters, so 2MASS photometry is used here.}} $K_S$ magnitude of the stars as the estimate of the intrinsic luminosity of the stars. As discussed in Section \ref{sec:photometry}, the distance modulus of \eso\ is only constrained by the turn-off region of the EFOSC2 CMD, whereas typically the HB luminosity would be a key anchor point. However, the effect of an error in the distance modulus on the metallicity estimate is small: for the \citet{Carrera2013} relation $d \feh/d \mathrm{K_S} \approx 0.15$~dex/magnitude.

Because there are no spectroscopic metallicity estimates for \eso\ with which to compare, it is prudent to repeat the reduction and analysis with other metal-poor clusters with well-established metallicities. In this work, we have used NGC4590, NGC7099, and NGC5053. These clusters were selected because a reasonable sample of their RGB stars had been observed with the AAOmega spectrograph with the same instrument configuration in the red camera as \eso. Their spectra were acquired from the AAT archive and reduced in the same way as the \eso\ spectra and their RGB member stars were identified using Pan-STARRS1 photometry \citep{Chambers2016}. The $(m-M)_V$ and \ebv\ collated in \citet{Harris1996} were used (along with $R_V=3.1,R_K=0.310$) to transform the 2MASS photometry of the stars to their absolute $K_S$ magnitudes. The radial velocities and metallicities were estimated for each star in the same manner as described in Section \ref{sec:rv_members}. Figure \ref{fig:violin} and Table \ref{table:cluster_aaomega} compare the literature metallicities of these clusters with the values estimated in this work. For these very metal-poor clusters the median CaT metallicity of the stars is different from the literature values by $<0.07$~dex.

While we identified 13 members of \eso, three of them (the two faintest stars in the top panel of Figure \ref{fig:phot_ew_feh}) are not in the 2MASS catalogue, so we only have metallicities for ten of them. From the ten stars, we estimate that \eso\ has a metallicity of $-2.48\substack{+0.06 \\ -0.11}$ (the median metallicity of the stars, with the 16th and 84th percentile ranges). This would make it the most metal-poor globular cluster of the Milky Way.

According to \citet{Harris1996} there are only two clusters with $\feh<-2.3$: NGC6341 ($\feh=-2.31$) and NGC7078 ($-2.37$). In both cases there are ranges of reported metallicities in the literature. For NGC6341, values range from $-2.12$ \citep{Beers1990} to $-2.5$ \citep{Peterson1993}, with most about $-2.3$ \citep{Armosky1994,Shetrone1996,Carretta2009}. Similarly, for NGC7078 there is a grouping of metallicity estimates around $-2.3>\feh>-2.4$ \citep{Armosky1994,Sneden1997,Kirby2008,Carretta2009}, though there is a couple of very metal-poor values reported: $-2.64$ \citep{Preston2006} and $\feh=-2.56$ \citep{Sobeck2011}. High resolution spectroscopic follow-up is required to confirm that \eso\ is the most metal-poor globular cluster.

\section{Are there unobserved cluster members?}\label{sec:photom_feh}

\begin{figure}
    \includegraphics[width=\columnwidth]{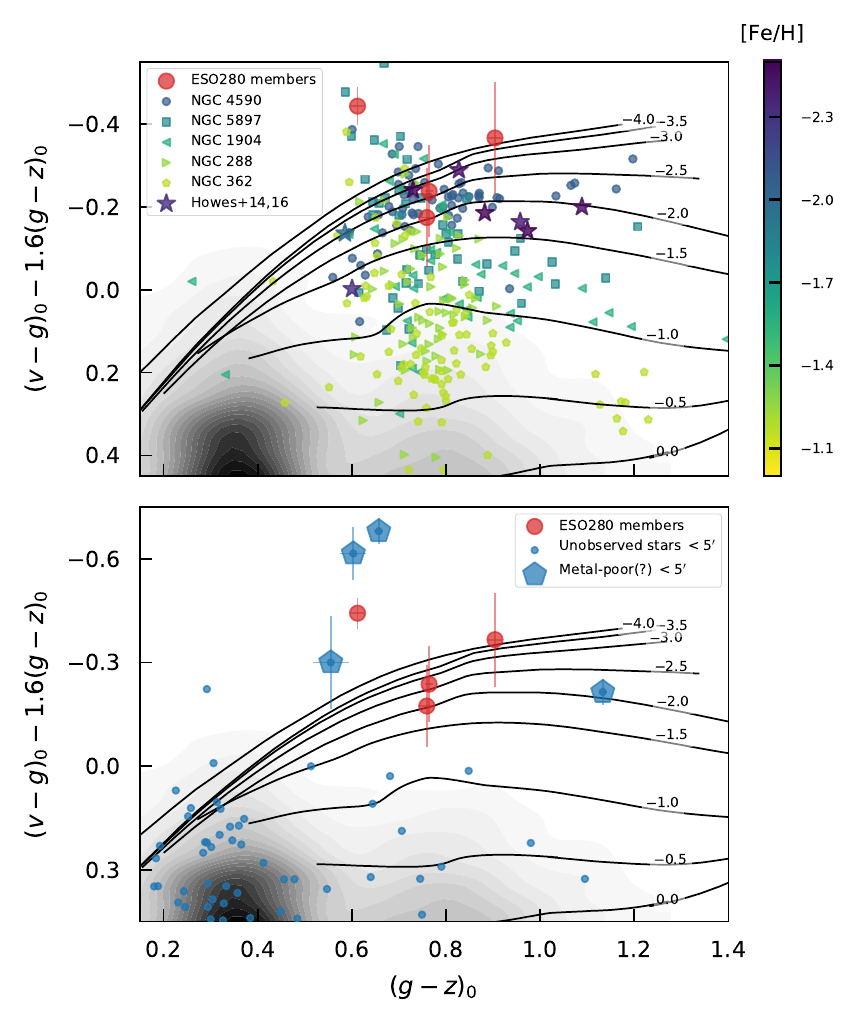}
    \caption{The $(v-g)_0-1.6(g-z)_0$ colour index is sensitive to the metallicity of the star and can be used to search for very metal-poor stars. The background contours give the distribution of 6000 random stars within 5 degrees of \eso, and {the 13-Gyr RGB isochrones from Dartmouth} are shown for a range of metallicities. The large red dots are the four \eso\ members with $v$ photometry. Top: the colour indices for stars from five globular clusters of a range of metallicities, and very metal-poor stars found by \citet{Howes2014, Howes2016}. This demonstrates the metallicity sensitivity of the index. Bottom: A search for possible members of \eso\ that were not observed spectroscopically. There are four possible very metal-poor giants with colour indices $<-0.2$ and $(g-z)_0>0.5$ (the dwarf/giant cut). These stars are possible members of \eso, especially when it is placed on the CMD (Figure \ref{fig:cmds_sm}).}
    \label{fig:sm_vgi}
\end{figure}

\begin{figure}
    \includegraphics[width=\columnwidth]{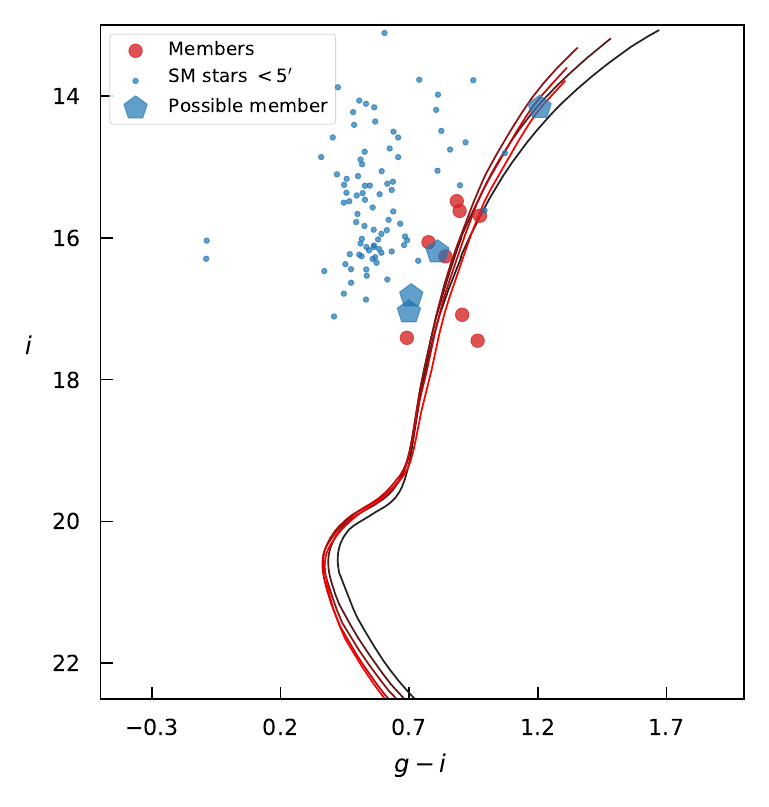}
    \caption{There are four unobserved, possible metal-poor star within 5~arcmin of \eso\ identified in Figure \ref{fig:sm_vgi} from their SkyMapper colour indices. They are consistent with the giant branch of the isochrones previously fitted to the CMD but spectroscopic follow-up is required to associate them with the cluster.}
    \label{fig:cmds_sm}
\end{figure}

The very low metallicity of \eso\ allows us to search for possible members of the cluster using the photometry from the SkyMapper survey. The SkyMapper telescope is a 1.3-m telescope at Siding Spring Observatory that is undertaking a multi-epoch photometric survey of the whole of the southern sky in six photometric bands: $uvgriz$ \citep{Keller2007}. The filter set of SkyMapper is similar to that of Sloan Digital Sky Survey \citep{Gunn1998} and Pan-STARRS \citep{Tonry2012}, but with some key differences. Relevant to this work is the addition of a narrow $v$ filter centred at 384~nm that is similar to the DDO 38 band and is very metallicity-sensitive\footnote{As noted in Section \ref{sec:photometry}, the $v$ filter bandpass is also gravity sensitive.}, especially at low metallicities \citep{Bessell2011}, and has been used in a number of works to identify very metal-poor candidate stars \citep[e.g.,][]{Keller2014,Howes2014,Howes2016,Jacobson2015}. Since the CaT-derived metallicity has confirmed that \eso\ is very metal-poor, we can use the SkyMapper DR1.1 photometry to search for possible members that were not part of our spectroscopic observing campaign.

A metallicity-sensitive colour index was constructed by combining the photometry from the $v,g,z$ filters: $(v-g)_0-1.6(g-z)_0$. The photometry of a given star was corrected for interstellar reddening using the \ebv\ computed by the SkyMapper team using the \citet{Schlegel1998} extinction maps, and bandpass absorption coefficients for the SkyMapper filters: $R_v=4.026,R_g=2.986,R_z=1.206$ \citep{Wolf2017}. Although we found a lower \ebv\ for the cluster than the \citet{Schlegel1998} value (Section \ref{sec:photometry}), we adopt their value here for consistency when comparing different regions of the sky. The members of \eso\ observed are near the faint limit of SkyMapper DR1.1, with only 8 of the 13 members catalogued, and only four of these eight have $v$ magnitudes required for the colour index.

To demonstrate the metallicity dependence of the colour index, in the top panel of Figure \ref{fig:sm_vgi} the indices of the \eso\ member stars are compared to stars with known metallicities. We also plot {13-Gyr, $[\alpha/\textrm{Fe}]=+0.4$ RGB isochrones from the Dartmouth Stellar Evolution Database \citep[][2012 version]{Dotter2008}} for a range of metallicities. The stars of known metallicity come from the globular clusters NGC4590 ($\feh=-2.06$), NGC 5897 ($-1.80$), NGC 1904 ($-1.54$) NGC 288 ($-1.24$), \& NGC 362 ($-1.16$), and seven of the very metal-poor stars ($-1.99<\feh<-2.80$) observed by \citet{Howes2014,Howes2016}. There is a clear trend of metallicity with the colour index and the \eso\ members have indices consistent with the low metallicity estimated from their CaT lines.

In the bottom panel of Figure \ref{fig:sm_vgi} we search for possible members of \eso\ that were not observed with AAOmega. Plotted are all the SkyMapper DR1.1 sources within 5~armin of \eso, with most of the stars having colours consistent with just being solar metallicity dwarf and giant field stars. But there are four stars with a colour index $<-0.2$ and with $(g-z)_0>0.5$ (the dwarf/giant cut): their SkyMapper DR1.1 designations are 180928.78-462616.9, 180921.02-462444.3, 180858.99-462131.8, and 180900.69-462737.6. On Figure \ref{fig:cmds_sm} these are placed on the SkyMapper CMD of \eso, and they do match the previously fitted isochrones. Spectroscopic follow-up is required to confirm these stars as a cluster members.

\section{Concluding remarks}\label{sec:discussion}

In this work, we have presented the first spectroscopic results for the very metal-poor globular \eso. Using new photometry of the cluster, we have confirmed the results from \citet{Ortolani2000} that \eso\ is a located $15.2\pm2.1$~kpc from the Galactic centre and with a heliocentric distance of $22.9\pm2.1$~kpc. This possibly places the cluster is the halo of the Galaxy, but, although we now have a radial velocity of $93.1\pm0.2$~km\,s$^{-1}$ for the cluster, we lack reliable proper motions to be certain of the cluster's orbit.

Overall, two findings from this work stand out: the cluster is more metal-poor than any other Milky Way globular cluster, with a metallicity estimated from the calcium triplet lines of $-2.48\substack{+0.06 \\ -0.11}$; and the photometry suggests that the cluster lacks a well-populated horizontal branch.

There is an apparent floor in the metallicity distribution function for GCs at $\feh=-2.5$; both in the Milky Way \citep{Harris1996} and Local Group dwarf galaxies \citep[e.g.,][]{Larsen2012}. The fact that no cluster has been observed below this limit has led to speculation that proto-galaxies must achieve a certain mass and/or stage in chemical evolution to form GCs \citep{Kruijssen2015}. \eso\ sits right at the apparent limit of the allowed cluster metallicity range. This implies that it formed in a primitive galaxy environment, making it a key relic of high-redshift star formation.

The possible lack of an HB in \eso\ is quite unexpected. Although GCs exhibit a wide range of HB morphology independent of their metallicity \citep[the ``second-parameter problem''; see e.g.,][]{Milone2014}, the clusters with no HB --- e.g., AM-4, E3, and Palomar 1 --- are young and metal-rich. {But \eso\ is old and metal-poor. There are old, metal-poor clusters in external galaxies with sparse HBs \cite[e.g., Fornax 1;][]{Buonanno1998,DAntona2013}, but their HB are still obvious.}

{It is informative to estimate how many HB stars should be expected for a cluster of \eso's mass. We estimate that the cluster velocity dispersion of $\sigma_v = 1.5\pm0.1$~km\,s$^{-1}$ from creating a 10000 iterations of the radial velocities for each star (excluding Star 1, 4, 12, 21, and 22 which have velocity errors larger than 1~km\,s$^{-1}$), with each iteration adjusted by the velocity error multiplied by a number drawn from a normal distribution. Using a half-light radius of $2.3$~pc, this equates to a dynamical mass of $(12\pm2
)\times10^3$~M$_\textrm{\sun}$ \citep[per the relations of e.g.,][]{Strader2009} --- within the normal bounds for globular clusters. For the other metal-poor clusters, there are between 20--60 blue HB stars observed, and these clusters have a mass range between 60000-160000 solar masses \citep{McLaughlin2005}, which equates to one HB star per 1000--5000 solar masses of the cluster. So for \eso, we might therefore expect on the order of 2--14 HB stars.} Is there some unexpected effect of cluster chemical evolution at this low metallicity that affects late stages of stellar evolution \citep[e.g.,][]{Campbell2012, MacLean2016a}? Are the most evolved stars in \eso\ preferentially being lost from the cluster?

The mass estimate does imply a high mass-to-light ratio. \citet{Bonatto2008} estimated an absolute magnitude for \eso\ of $M_V=-4.9\pm0.3$, which would mean a mass-to-light ratio $M/L_V=1.6$. This can be compared to typical values for globular clusters for the Milky Way and M31 where are 1--2 \citep[e.g.,][]{Strader2009,Kimmig2015}. So \eso\ does have a typical mass-to-light ratio despite the sparsely populated red giant branch and a non-existent HB.

There are several follow up observations that would hopefully clarify our view of \eso. Detailed chemical abundance studies from high-resolution spectroscopy will allow us to: place \eso\ in context with well-studied GCs; search it for multiple stellar populations; investigate the process of cluster self-enrichment in this unexplored metallicity regime. A key region of the CMD to investigate are the stars that possibly form a red HB of the cluster. Further spectroscopic observations of the cluster will allow us to confirm if the four stars identified from SkyMapper's colour index is indeed a cluster member. It would be extremely helpful to search for further extra-tidal stars beyond the one possible star identified in this work. If ongoing mass loss is responsible for the small population of evolved stars of the cluster, these observations will find  stars in the process of escaping, and if there is not a significant number of stars being lost from the cluster, then the CMD morphology might be a feature of stellar evolution at extremely low metallicity.

Two near-future survey releases should include helpful data for clarifying our picture of \eso: \textit{Gaia} DR2 and the SkyMapper Main Survey. \textit{Gaia} DR2 is planned for release on 2018 April 25 and will likely include parallaxes, proper motions, and a blue \& red photometry for all of our observed members. Simulations by \citet{Pancino2017b} have shown that this photometry will clearly separate the HB from the RGB for globular clusters. We will be able to use parallaxes to separate some of the field star population from the cluster stars. The SkyMapper Main Survey will increase the faint limit by 3--4 magnitudes over the SkyMapper DR1.1 data used here, which should definitely include any possible blue HB and also provide the gravity sensitive $u-v$ colour of the stars. With a planned faint limit of $g=21.7$, this should just include the turn-off region of the CMD.

\section*{Acknowledgements}
JDS wishes to thank Sarah Martell, Daniel Zucker, and Peter Cottrell for their helpful input into this work, and the referee for their helpful comments and suggestions that improved this manuscript.

The data in this paper were based on observations obtained at the Australian Astronomical Observatory as part of programmes A/2011B/20 (NGC1904), A/2010B/18 (NGC2298), S/2016A/13 (\eso). We are grateful to AAO Director Warrick Couch for awarding us Director's Discretionary Time on the AAT which expanded our dataset. We acknowledge the traditional owners of the land on which the AAT stands, the Gamilaraay people, and pay our respects to elders past and present.

Based on observations made with ESO Telescopes at the La Silla or Paranal Observatories under programme ID(s) 095.D-0037(A), 179.A-2010(B), 179.A-2010(C), 179.A-2010(D), 179.A-2010(E), 179.A-2010(F), 179.A-2010(G), 179.A-2010(H), 179.A-2010(I), 179.A-2010(J), 179.A-2010(K), 179.A-2010(L), 179.A-2010(M).

The following software and programming languages made this research possible: \textsc{esorex}, the ESO Recipe Execution Tool; \textsc{2dfdr} \citep[v6.46;][]{AAOSoftwareTeam2015}, the 2dF Data Reduction software; Python (v3.6.2); \textsc{astropy} \citep[v3.0;][]{Robitaille2013,TheAstropyCollaboration2018}, a community-developed core Python package for Astronomy; \textsc{pandas} \citep[v0.22;][]{McKinney2010}; \textsc{seaborn} \citep[v0.8.1;][]{Waskom2017}; Tool for OPerations on Catalogues And Tables \citep[\textsc{topcat}, v4.5;][]{Taylor2005}.

The national facility capability for SkyMapper has been funded through ARC LIEF grant LE130100104 from the Australian Research Council, awarded to the University of Sydney, the Australian National University, Swinburne University of Technology, the University of Queensland, the University of Western Australia, the University of Melbourne, Curtin University of Technology, Monash University and the Australian Astronomical Observatory. SkyMapper is owned and operated by The Australian National University's Research School of Astronomy and Astrophysics. The survey data were processed and provided by the SkyMapper Team at ANU. The SkyMapper node of the All-Sky Virtual Observatory (ASVO) is hosted at the National Computational Infrastructure (NCI). Development and support the SkyMapper node of the ASVO has been funded in part by Astronomy Australia Limited (AAL) and the Australian Government through the Commonwealth's Education Investment Fund (EIF) and National Collaborative Research Infrastructure Strategy (NCRIS), particularly the National eResearch Collaboration Tools and Resources (NeCTAR) and the Australian National Data Service Projects (ANDS).

The Pan-STARRS1 Surveys (PS1) and the PS1 public science archive have been made possible through contributions by the Institute for Astronomy, the University of Hawaii, the Pan-STARRS Project Office, the Max-Planck Society and its participating institutes, the Max Planck Institute for Astronomy, Heidelberg and the Max Planck Institute for Extraterrestrial Physics, Garching, The Johns Hopkins University, Durham University, the University of Edinburgh, the Queen's University Belfast, the Harvard-Smithsonian Center for Astrophysics, the Las Cumbres Observatory Global Telescope Network Incorporated, the National Central University of Taiwan, the Space Telescope Science Institute, the National Aeronautics and Space Administration under Grant No. NNX08AR22G issued through the Planetary Science Division of the NASA Science Mission Directorate, the National Science Foundation Grant No. AST-1238877, the University of Maryland, Eotvos Lorand University (ELTE), the Los Alamos National Laboratory, and the Gordon and Betty Moore Foundation.

This publication makes use of data products from the Two Micron All Sky Survey, which is a joint project of the University of Massachusetts and the Infrared Processing and Analysis Center/California Institute of Technology, funded by the National Aeronautics and Space Administration and the National Science Foundation.

This research has made use of the SVO Filter Profile Service (\url{http://svo2.cab.inta-csic.es/theory/fps/}) supported from the Spanish MINECO through grant AyA2014-55216



%



\bsp    
\label{lastpage}
\end{document}